\documentstyle[pra,aps,twocolumn,epsfig]{revtex}

\newcommand{\eins}{\mbox{$1 \hspace{-1.0mm}  {\bf l}$}}
\newcommand{\be}{\begin{equation}}
\newcommand{\ee}{\end{equation}}
\newcommand{\bea}{\begin{eqnarray}}
\newcommand{\eea}{\end{eqnarray}}

\newcommand{\ket}[1]{ | \, #1  \rangle}
\newcommand{\bra}[1]{ \langle #1 \,  |}

\begin{document}

\title{Reversible quantum teleportation  in an optical lattice}

\author{Luis Santos and Dagmar Bru\ss }

\address{Institut f\"ur Theoretische Physik, Universit\"at Hannover,
 Appelstr. 2, D--30167 Hannover,
Germany}

\maketitle

\begin{abstract}
We propose a protocol, based
on entanglement procedures recently suggested by [D. Jaksch {\it et al.}, Phys.
Rev. Lett. {\bf 82}, 1975 (1999)], which allows the teleportation 
of an unknown state of a neutral atom in an optical lattice to another atom in
another site of the lattice, without any irreversible detection.

\end{abstract}

\section{Introduction}
\label{sec:intro6}

The characterization, creation, control and manipulation of quantum
entanglement \cite{Einstein35,Bellbook} constitutes the basis of  the
 fast--developing research field 
of quantum information.  
An
entangled state of two or more  
 \cite{Greenberger89} particles can be
intuitively understood as the situation in which the state of one particle
cannot be determined  independently from the state of the others. Of course, the
concept deserves more technical definitions and quantifications, but these are
beyond the scope of this paper. 

	Quantum computing \cite{Ekert96}, quantum cryptography \cite{Ekert92},
and other interesting phenomena, constitute excellent examples of the
extraordinary potential of quantum entanglement. Among these phenomena,
quantum teleportation \cite{Bennet93} is perhaps one of the most striking
ones. Quantum teleportation consists in the transport of a quantum state from
one particle to another in a disembodied way. Following the theoretical
proposal of Bennett et al \cite{Bennet93}, quantum teleportation has been
recently demonstrated experimentally \cite{Bowmeester97}. This teleportation
scheme involves a  Bell-measurement \cite{foot1}, and therefore it
is performed in an irreversible way. However, it has been shown
\cite{Braunstein96,Nielsen97} that quantum teleportation can be performed in a
reversible way, i.e. without any irreversible detection. In order to obtain
such unitary teleportation it is necessary to consider the detector 
as a quantum mechanical object, the state of which  is not 
read out.
 In this paper, we show how this idea
can be implemented in a particular physical situation.

	Several physical systems have been proposed, in which entanglement can
be created and manipulated, as for example cavity QED \cite{Turchette95},
photons \cite{Aspect82} and ion--traps \cite{Cirac95}. Recently it has
been proposed that neutral atoms in an optical lattice can be entangled in a
controlled way by using cold collisions between them \cite{Jaksch99}. It has
been shown  \cite{Briegel99} how one can use this 
entanglement mechanism to create GHZ-states \cite{Greenberger89}, or implement
parallel quantum computing and quantum error correction \cite{Shor95}. We
 show in the present paper, that in the framework of this particular
entanglement procedure it is possible to implement a reversible teleportation
protocol.

	The structure of the paper is as follows. In Sec.\ \ref{sec:tel} we
review the irreversible and reversible teleportation schemes, and
also present the abstract formulation of our teleportation proposal. In Sec.\
\ref{sec:opt}, we briefly review the entanglement procedure of
\cite{Jaksch99,Briegel99}. In Sec.\ \ref{sec:exa}, we consider 
explicitly the most
simple case of teleportation with 
just three lattice sites, whereas in Sec.\
\ref{sec:gen} we analyze the case of an arbitrary number of sites. 
We finalize in Sec.\ \ref{sec:conclu} with some conclusions.

\section{Teleportation protocols}
\label{sec:tel}

In this section, we review briefly the general ideas behind  the
teleportation schemes both with and without irreversible measurements. 
For a more
detailed discussion we refer to Refs.\ \cite{Bennet93,Braunstein96,Nielsen97}. At the
end of the section we present the abstract formalism of our teleportation
scheme. In this section, we follow the notation of Ref.\
\cite{Braunstein96}. 

\subsection{Irreversible teleportation}
\label{sec:irr}
We are considering three two--level systems, 
denoted by index 1,2 and 3, each
 with basis states $\{|0\rangle , |1\rangle\}$. 
Initially  particle 1 is in an unknown state $|\phi\rangle$, whereas
particles 2 and 3 are in a maximally entangled state 
$|\phi^+\rangle= (|0\rangle |0\rangle + |1\rangle |1\rangle )/\sqrt{2}$. 
Particles 1 and 2 are together at one place, and particle three is at a
different place.
The
teleportation scheme can be well understood using the decomposition
\cite{Bennet93}:
\begin{equation}
\ket{\phi_{in}}=|\phi\rangle_1 |\phi^+\rangle_{23}=\frac{1}{2}\sum_{J=0}^3
|\psi^{(J)}\rangle_{12} U_3^{(J)\dag}|\phi\rangle_3,
\label{normal}
\end{equation}
where $|\psi^{(J)}\rangle$ is the entangled basis for two qubits,
\bea
|\psi^{(J)}\rangle & = & \sum_{l=0}^1 \exp{(\pi i l n)}\ket{l}
           \ket{(l+m) \text{mod}\, 2}/\sqrt{2} \ \ , \nonumber \\
  U^{(J)}  & = &  \sum_{k=0}^1 \exp{(\pi i k n)}\ket{k}
           \bra{(k+m) \text{mod}\, 2}        
\eea
with
\be
J=n\cdot 2 +m \ \ , \ \ \ \ \text{i.e.} \ n=J\text{div} 2; \ m= J\text{mod} 2\ .
\ee
Explicitly this corresponds to
$|\psi^{(0)}\rangle\equiv |\phi^{+}\rangle$, 
$|\psi^{(1)}\rangle\equiv |\psi^{+}\rangle$, 
$|\psi^{(2)}\rangle\equiv |\phi^{-}\rangle$, 
$|\psi^{(3)}\rangle\equiv |\psi^{-}\rangle$,  the well--known Bell states
\cite{foot1}, and $ U^{(0)}\equiv \eins$, $ U^{(1)}\equiv
\sigma_1$,  $U^{(2)}\equiv\sigma_3$, $ U^{(3)}\equiv i
\sigma_2$, where $\sigma_j$ are the Pauli matrices. 
\par
The teleportation scheme
\cite{Bennet93} works as follows: first 
 a joint measurement of the state of particles $1$
and $2$, i.e. a Bell measurement, is performed; then, using a
classical channel, the result $J$ of the measurement is sent to 
the other site and,
using the  value $J$, the appropriate unitary
transformation $ U_3^{(J)}$ is performed on particle $3$ to transform the
state of  particle $3$ into $|\phi\rangle$. As we observe, this procedure is
clearly irreversible, because a measurement of the joint state of $1$ and $2$
is necessary.
\par In the following a lower index 1,2,3,... for a unitary transformation refers
to the particle to which the transformation is applied, and a lower index 
$a,b,c$ refers to the sequence of transformations.

\subsection{Reversible teleportation}
\label{Rev}
However, the previous scheme is not the only one which allows to teletransport
an unknown state. In particular, there is a reversible way to perform such
a task \cite{Braunstein96,Nielsen97}. Let us assume an auxiliary particle $A$
(which in the following we shall call ancilla), which is a
four--level system with basis states $\{\ket{0},\ket{1},
\ket{2},\ket{3}\}$.
The initial state of the system (\ref{normal}) becomes now:
\begin{equation}
|\phi_{in}\rangle=|0\rangle_A|\phi\rangle_1 |\phi^+\rangle_{23},
\label{newini}
\end{equation}
The reversible
teleportation scheme works as follows. First, we perform the following
unitary operation on the initial state:
\begin{equation}
 U_a=\sum_{J=0}^3|\psi^{(J)}\rangle_{12} O_A^{(J)}\
\langle\psi^{(J)}|_{12}, \label{U1}
\end{equation}
where $ O_A^{(0)}\equiv \eins$ and 
$ O_A^{(J)}\equiv |0\rangle\langle J |+ |J\rangle\langle 0 |+\sum_{k\not=
0,J}^3 |k\rangle\langle k|$, for $0<J\leq 3$. It is easy to observe that the
state of the system becomes:
\begin{equation}
 U_a |\phi_{in}\rangle=\frac{1}{2}\sum_{J=0}^3 |J\rangle_A|\psi^{(J)}\rangle_{12}
U_3^{(J)\dag}|\phi\rangle_3 .
\label{U1phi0}
\end{equation}
A second step consists in transporting the particle $A$ near  particle $3$,
and performing a unitary transformation of the form:
\begin{equation}
 U_b=\sum_{J=0}^3 |J\rangle_A  U_3^{(J)}\ \langle J|_A,
\label{U2}
\end{equation}
i.e. a unitary transformation on $3$ conditional to the value of the state of
the ancilla $A$. After performing operation $ U_b$ the system becomes
\begin{equation}
 U_b U_a |\phi_{in}\rangle=\frac{1}{2}|\phi\rangle_3\sum_{J=0}^3
|J\rangle_A|\psi^{(J)}\rangle_{12}
\label{U1U2}
\end{equation}
and therefore the state $|\phi\rangle$ has been transported into $3$, whatever
the final value of $1$, $2$ and $A$. Eventually, if $A$ were transported back
near $1$ and $2$, one could perform a unitary operation to restore $1$, $2$
and $A$ to their original states.

\subsection{Our teleportation scheme}
In this section, we present the general idea behind our
reversible teleportation scheme, in order to observe the
similarities and differences in comparison with
 the scheme presented previously. In the
following we consider 
three two-level systems
$1$, $2$, $3$. Note that, as we show in Sec.\ \ref{sec:gen}, the method
can be generalized to an  arbitrary number of particles. 
The lowest non-trivial number of particles is three and therefore used
to explain our scheme.
Due to
the restrictions of the physical model that we employ in Sec.\ \ref{sec:opt},
all particles, including the ancillas, are in our case qubits, i.e. have just
two states $\{|0\rangle,|1\rangle\}$ (in the case of the ancillas we shall
consider a different state $|2\rangle$ instead of $|1\rangle$).  In order to
perform the teleportation, though, we need two bits of information  and
therefore two ancillas $A_1$ and $A_2$. We assume that the initial state of
the system is of the form 
\begin{equation}
|\phi_{in}\rangle
=|0\rangle_{A_1}|0\rangle_{A_2}|\phi\rangle_1|0\rangle_2|0\rangle_3.
\label{ini2}
\end{equation}
As one can observe, this is a  difference with respect to the initial state
considered in expression (\ref{normal}): the particles $2$ and
$3$ are not  initially entangled, and as we will see later they will 
become entangled with each other as well as  with particle 1 during our
transformations.

The first step of our teleportation scheme consists in performing a unitary
transformation $V_a$ acting on the $6$-dimensional space of the three
particles $1$, $2$ and $3$, in such a way that the state of the system becomes
\begin{equation}
V_a\ket{\phi_{in}} = \sum_{J=0}^3 |0\rangle_{A_1}|0\rangle_{A_2}
      |\tilde{\psi}^{(J)}\rangle_{12}
{\tilde U}_3^{(J)\dag}|\phi\rangle_3 ,
\label{Lphi0}
\end{equation}
where the exact definitions of $|\tilde{\psi}^{(J)}\rangle_{12}$, 
${\tilde U}_3^{(J)}$, and $V_a$ will be presented in Sec.\
\ref{sec:exa}. As we observe, the operation $V_a$ entangles the particles
$1$, $2$ and $3$. The next step of the teleportation scheme is to perform a
unitary operation of the form:
\begin{equation}
 V_b\equiv\sum_{J=0}^3|\tilde{\psi}^{(J)}\rangle_{12}
O_{A_1A_2}^{(J)}\ \langle\tilde{\psi}^{(J)}|_{12},
\label{B1}
\end{equation}
where $ O_{A_1A_2}^{(J)}=\sigma_{1,A_1}^{J\text{mod}2}
\sigma_{1,A_2}^{J\text{div}2}$ (our notation is $\sigma_i^0=\eins$ and 
$\sigma_i^1=\sigma_i$).
After applying $ V_b$ the state of the system takes the form
\begin{equation}
V_bV_a\ket{\phi_{in}} =\sum_{J=0}^3|\tilde{\psi}'^{(J)}\rangle_{A_1A_2}
|\tilde{\psi}^{(J)}\rangle_{12}
U_3^{(J)\dag}|\phi\rangle_3 
\label{B1Lphi0}
\end{equation}
where the exact form of $|\tilde{\psi}'^{(J)}\rangle_{A_1A_2}$ is presented in
Sec.\ \ref{sec:exa}. This step can be called the ``reading of the states by the
ancillas". Once this is done, we perform a unitary operation of the form:
\begin{equation}
 V_c\equiv\sum_{J=0}^3|\tilde{\psi}'^{(J)}\rangle_{A_1A_2}
 R_3^{(J)}\ \langle\tilde{\psi}'^{(J)}|_{A_1A_2},
\label{C1}
\end{equation}
where $ R_3^{(J)}{\tilde U}_3^{(J)\dag}=c_J{\tilde
U}_3^{(0)\dag}$. Here the coefficients $c_J$ can be $\pm 1$, and are in
detail calculated in Sec.\ \ref{sec:exa}. Then, the state of the system
becomes 
\begin{equation}
 V_cV_bV_a\ket{\phi_{in}} ={\tilde U}_3^{(0)\dag}|\phi\rangle_3 \otimes 
\sum_{J=0}^3 c_J |\tilde{\psi}'^{(J)}\rangle_{A_1A_2}
|\tilde{\psi}^{(J)}\rangle_{12},
\end{equation}
where we have moved the state of $3$ out of the sum to stress that it
 is now independent of the state of $1$, $2$ and the ancillas. As a
final step, we just need to perform the unitary operation  
$ {\tilde U}_3^{(0)}$ to
conclude the teleportation. The ancillas and $1$ and $2$ remain in an
entangled state. Finally, but this is not necessary for the teleportation, if
we perform again $ V_b$ the ancillas are brought back to their original
state $|0\rangle_{A_1}|0\rangle_{A_2}$.

\section{Quantum entanglement of atoms in optical lattices}
\label{sec:opt}
In this section we present the physical system in which we shall implement our
reversible teleportation scheme, briefly reviewing
Refs. \cite{Jaksch99,Briegel99}. Let us consider a collection of bosonic
neutral atoms occupying the sites of an optical lattice. In order to perform
the necessary quantum logical operations one has to be able to fill the lattice
wells with exactly one particle each. This can be achieved 
- at present only theoretically - 
by loading the lattice from a
Bose--Einstein Condensate (BEC), and inducing at sufficiently low temperatures
a phase transition from the superfluid BEC phase into a Mott insulator phase,
by increasing the ratio between the interaction energy inside each well and
the tunneling rate between the wells, as predicted by the Bose--Hubbard model
\cite{Jaksch98}. In addition to single--atom operations \cite{foot1b} two
basic two--atom operations can be perfomed within this physical scheme,
based on cold collisions between the atoms.

\subsection{Shift operation}
\label{sec:shi}

The two internal states of the atoms carrying the
quantum information are called $\{|0\rangle,|1\rangle\}$. As
shown in detail in Ref.\ \cite{Briegel99}, by properly arranging the detuning
and polarization of the lasers which form the lattice, it can be achieved
that each of the internal atomic levels can see a different potential. In
particular, by changing the dephasing between the circularly polarized waves
$\sigma^{\pm}$ which form the lattice,  the potentials for the different
internal levels move in opposite directions. Let us suppose that the
lattice of $|0\rangle$ moves to the right, while the lattice of $|1\rangle$
moves to the left. Two neighbour atoms in the lattice
 occupy sites $j$ and $j+1$
 (our numbering is from left to right). 
It is clear that, using a lattice displacement, the neighbour atoms can only
undergo a collision if the atom at $j$ is in $|0\rangle$ and the one at $j+1$
is in $|1\rangle$. Any other situation prevents the atoms 
from approaching each other.
When the two particles are put in contact,  they can interact via s--wave
scattering. As a result, a collisional phase appears, which can be controlled
by basically changing the interaction time. In particular, we shall choose
this collisional phase to be $\pi$. After the desired time the lattice is
brought back to its original position \cite{foot2}. We shall call this
operation {\em shift operation}, following the notation of \cite{Briegel99},
and it can be summarized in the following table: \begin{eqnarray}
|0\rangle_j|0\rangle_{j+1}&\rightarrow& |0\rangle_j|0\rangle_{j+1},\nonumber\\
 |0\rangle_j|1\rangle_{j+1}&\rightarrow&-|0\rangle_j|1\rangle_{j+1},\nonumber\\
 |1\rangle_j|0\rangle_{j+1}&\rightarrow&
|1\rangle_j|0\rangle_{j+1},\nonumber\\  |1\rangle_j|1\rangle_{j+1}&\rightarrow
&|1\rangle_j|1\rangle_{j+1} \ .  \label{tableshift} \end{eqnarray}

\subsection{Sweep operation}
\label{secswe}

Let us assume that the atoms have a third atomic level $|2\rangle$, which 
 can be displaced like the levels $|0\rangle$ and $|1\rangle$ by using the
corresponding transport lattice. The operation is basically like the previous
one, but now only those atoms in level $|2\rangle$ are going to participate.
In particular, we are going to consider that just the ancilla 
is excited into
the level $|2\rangle$. The interaction of the ancilla with an atom in the site
$j$  can be designed (following the same arguments as
above) in such a way that\cite{foot3}:
\begin{eqnarray}
|2\rangle_A|0\rangle_j\rightarrow |2\rangle_A|0\rangle_j \nonumber \\
|2\rangle_A|1\rangle_j\rightarrow -|2\rangle_A|1\rangle_j
\label{tablesweep}
\end{eqnarray}
Following \cite{Briegel99}, we shall call this operation {\em sweep
operation}. By varying the speed by which the lattice of the state $|2\rangle$
is moved during the sweep operation, it is possible to act on a particular site
of the lattice, even when the ancilla crosses through other sites in the
lattice, in particular because the collisional time can
be designed in such a way that for undesired sites the collisional phase is a
multiple of $2\pi$ \cite{foot4}.

\section{Reversible teleportation protocol: three sites only}
\label{sec:exa}

Let us now show explicitly the transformations for our scheme of reversible
teleportation. In this section we explain the most simple
 case, in which we have three sites, each one
occupied by one atom. We shall call these atoms (from left to right in the
lattice) $1$, $2$ and $3$. We shall consider another two atoms $A_1$ and $A_2$,
 which will act as ancillas, and which are initially placed
sufficiently apart (at the left) from the sites $1$, $2$ and $3$; this
requirement is necessary to avoid that the unitary operations applied on the
site $1$, $2$ and $3$ could affect the ancillas, and vice versa.
We shall also
consider that both ancillas are separated by a distance larger than the
dimensions of the three sites $1$, $2$ and $3$; this requirement is necessary
to avoid that during the operation of one ancilla on the sites, the other
sites could be affected
by the other ancilla.  We shall discuss in Sec.\ \ref{sec:conclu} the case
in which only one ancilla is present. We begin with the initial state of the
system:

\begin{equation}
\ket{\phi_{in}}=
|0\rangle_{A_1}|0\rangle_{A_2}|\phi\rangle_1|0\rangle_2|0\rangle_3,
\label{ini3}
\end{equation}
where $|\phi\rangle=a|0\rangle+b|1\rangle$. Our objective is to transport this
state from particle $1$ to particle $3$, in a reversible way, using the
operations of Sec.\ \ref{sec:opt}. Our
teleportation scheme consists on the following three general steps:

\subsection{Creation of the entanglement between $1$, $2$ and $3$}
\label{sec:cre}
As a first step, we perform a Hadamard transform ($H$)
\begin{eqnarray}
|0\rangle\rightarrow\frac{1}{\sqrt{2}}(|0\rangle+|1\rangle), \nonumber \\
|1\rangle\rightarrow\frac{1}{\sqrt{2}}(|0\rangle-|1\rangle),
\label{Hadamard}
\end{eqnarray}
in each one of the sites $1$, $2$ and $3$ \cite{foot1b}. Then, we perform
a shift operation (of the lattices of $|0\rangle$ and $|1\rangle$), and after
that we perform another Hadamard transformation to all the sites. As a result
of these three operations the state of the system becomes:
\begin{equation}
V_a\ket{\phi_{in}}=
\frac{1}{2}|0\rangle_{A_1}|0\rangle_{A_2}\sum_{J=0}^{3}|\tilde{\psi}^{(J)}\rangle_{12}{\tilde
U}_3^{(J)\dag}|\phi\rangle_3,
\label{HLH}
\end{equation}
where $|\tilde{\psi}^{(J)}\rangle_{12}=|J\text{div}\, 2\rangle_1
|J\text{mod}\, 2\rangle_2$, and 
${\tilde U}^{(0)}=-i\sigma_2$, ${\tilde U}^{(1)}=\sigma_1$, 
${\tilde U}^{(2)}=-\sigma_3$, ${\tilde U}^{(3)}= \eins$. As we observe the
state of the three sites becomes now entangled.

\begin{figure}[ht]
\begin{center}\
\epsfxsize=7.5cm
\hspace{0mm}
\psfig{file=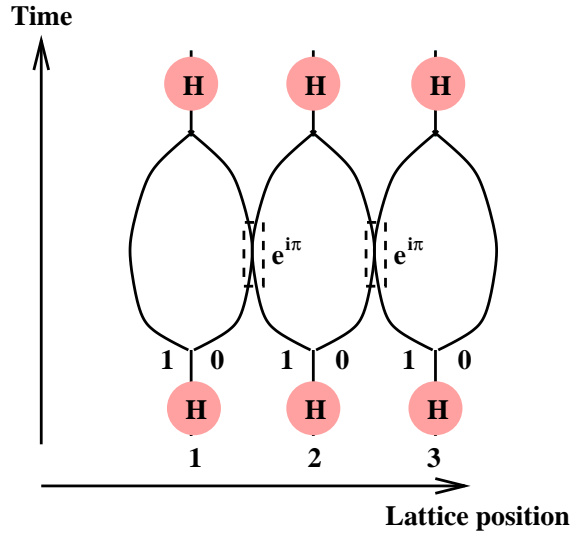,width=7.5cm}\\[0.1cm]
\caption{Creation of the entanglement between sites $1$, $2$ and $3$. We have
depicted the shift operation as a multiparticle interferometer following
Ref.\ [15]. Here $H$ denotes a Hadamard transform.}
\label{fig:tel1}
\end{center}
\end{figure}

\subsection{Reading of $1$ and $2$ using the ancillas}
\label{sec:ent}

First, a Hadamard transform is performed in each one of the ancillas (between
levels $|0\rangle$ and $|2\rangle$):
\begin{eqnarray}
H_AV_a\ket{\phi_{in}}
&=&\frac{1}{4}(|0\rangle_{A_1}+|2\rangle_{A_1})(|0\rangle_{A_2}+|2\rangle_{A_2})
\nonumber \\
&& \otimes \sum_{J=0}^{3}|\tilde{\psi}^{(J)}\rangle_{12}{\tilde
U}_3^{(J)\dag}|\phi\rangle_3.
\end{eqnarray}
Then, we perform a sweep to the right of the lattice of $|2\rangle$
until placing the ancilla $A_1$ in site $2$, and after that sweeping again to
the right we place the ancilla $A_2$ in site $1$. In both cases, the
interaction times are properly designed to obtain a sweep operation as
described in (\ref{tablesweep}). Finally we displace the lattice of
$|2\rangle$ back to its original position. The state of the system 
after this operation $\tilde{V_b}$ becomes:
\begin{eqnarray}
\tilde{V_b}H_AV_a\ket{\phi_{in}}
&=&\frac{1}{4}\sum_{J=0}^{3}
(|0\rangle_{A_1}+(-1)^{J\text{mod}2}|2\rangle_{A_1})\nonumber \\ && 
\ \ \ \ \ \
(|0\rangle_{A_2}+(-1)^{J\text{div}2}|2\rangle_{A_2}) \nonumber \\ && \otimes
|\tilde{\psi}^{(J)}\rangle_{12}{\tilde
U}_3^{(J)\dag}|\phi\rangle_3\  \ .
\end{eqnarray}
\begin{figure}[ht]
\begin{center}\
\epsfxsize=6.5cm
\hspace{0mm}
\psfig{file=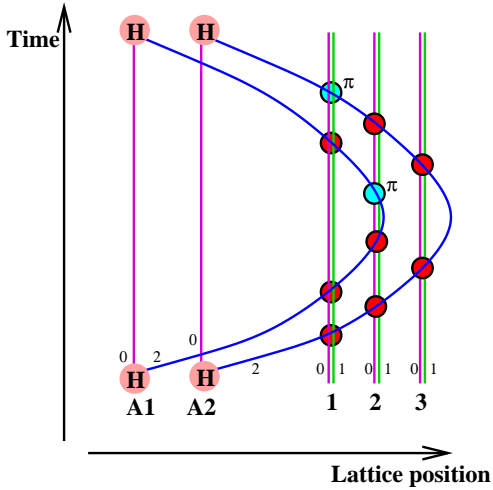,width=6.5cm}\\[0.1cm]
\caption{Reading of the state of the sites $1$ and $2$ using the
ancillas $A_1$ and $A_2$. $H$ denotes a Hadamard transform, and the
circles sweep operations. The sweep operations which introduce a phase $\pi$
as in (\ref{tablesweep}) are indicated. The rest are assumed to lead to a zero
phase.}
\label{fig:tel2}
\end{center}
\end{figure}
Performing a new Hadamard transform in both ancillas, the state of the system
becomes:
\begin{equation}
V_bV_a\ket{\phi_{in}}=
\frac{1}{2}\sum_{J=0}^{3}|\tilde{\psi}'^{(J)}\rangle_{A_1 A_2}
|\tilde{\psi}^{(J)}\rangle_{12}{\tilde
U}_3^{(J)\dag}|\phi\rangle_3,
\end{equation}
where $|\tilde{\psi}'^{(J)}\rangle_{A_1 A_2}=
|J\text{mod}\, 2\rangle_{A_1}|J\text{div}\, 2\rangle_{A_2}$
and $V_b=H_A\tilde{V_b}H_A$. 
Therefore, we have copied the state
of $1$ ($2$) into $A_2$ ($A_1$). Note that the joint operation $V_b$ is
equivalent to a sequence of two CNOT gates (i) with $2$ as control qubit and
$A_1$ as target; (ii) with $1$ as control qubit and $A_2$ as target.

\subsection{Teleportation}

Now, we are going to use the values of the ancillas to teleport the state
$|\phi\rangle$ into the site $3$. First of all, we perform a Hadamard transform
in the site $3$ (in principle this operation can be performed simultaneously in
the other two sites \cite{foot1b}, but since the other sites remain untouched
during this step, we just consider for simplicity that only the site $3$ is
affected by this  operation). Then, the lattice of $|2\rangle$ is
swept to the right until $|2\rangle_{A_2}$ is in contact with the site $3$ and
interacts following the rule of (\ref{tablesweep}). Let us call this operation
$\tilde V_c$. After that we perform again a Hadamard transform in the site
$3$. One calculates that the effect of these steps is to change
the state of the system into \begin{eqnarray}
&&\frac{1}{2}(|\tilde{\psi}'^{(0)}\rangle_{A_1 A_2}
|\tilde{\psi}^{(0)}\rangle_{12}-
|\tilde{\psi}'^{(2)}\rangle_{A_1 A_2}
|\tilde{\psi}^{(2)}\rangle_{12}){\tilde
U}_3^{(0)\dag}|\phi\rangle_3 \nonumber \\
&&+\frac{1}{2}(|\tilde{\psi}'^{(1)}\rangle_{A_1 A_2}
|\tilde{\psi}^{(1)}\rangle_{12}+
|\tilde{\psi}'^{(3)}\rangle_{A_1 A_2}
|\tilde{\psi}^{(3)}\rangle_{12}){\tilde
U}_3^{(1)\dag}|\phi\rangle_3.
\end{eqnarray}
\begin{figure}[ht]
\begin{center}\
\epsfxsize=7.5cm
\hspace{0mm}
\psfig{file=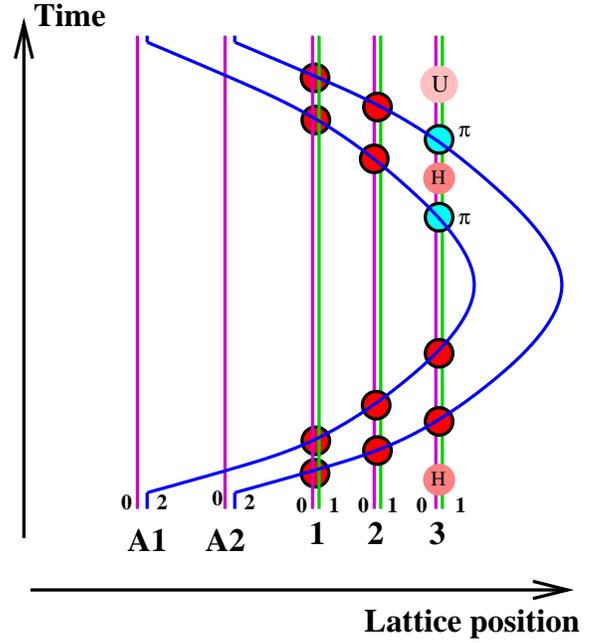,width=7.5cm}\\[0.1cm]
\caption{Using the values deposited in the ancillas during the reading
process, the teleportation is finalized using unitary one--atom operations in
site $3$, and one sweep of the lattice of the state $|2\rangle$. In the
graphic $H$ denotes the
Hadamard transform, $U\equiv {\tilde U}^{(0)}$, and the
circles sweep operations. The sweep operations which introduce a phase $\pi$
as in (\ref{tablesweep}) are indicated. The rest are assumed to lead to a zero
phase.}
\label{fig:tel3} 
\end{center} 
\end{figure}
After this, we move further to the right the lattice of $|2\rangle$ until 
$|2\rangle_{A_1}$ enters in contact with the site $3$, interacts with it 
following the rule of (\ref{tablesweep}), and then we sweep back the lattice of
$|2\rangle$ to its original position. Let us call this operation $\tilde
V'_c$. After the joint operation $V_c=\tilde V'_c H_A \tilde V_c H_A$, the
state of the system takes the form: 
\begin{eqnarray}
&&V_cV_bV_a\ket{\phi_{in}}=\nonumber \\
&&\frac{1}{2}(|\tilde{\psi}'^{(0)}\rangle_{A_1 A_2}
|\tilde{\psi}^{(0)}\rangle_{12}-
|\tilde{\psi}'^{(1)}\rangle_{A_1 A_2}
|\tilde{\psi}^{(1)}\rangle_{12}- \nonumber \\
&&|\tilde{\psi}'^{(2)}\rangle_{A_1 A_2}
|\tilde{\psi}^{(2)}\rangle_{12}-
|\tilde{\psi}'^{(3)}\rangle_{A_1 A_2}
|\tilde{\psi}^{(3)}\rangle_{12}){\tilde
U}_3^{(0)\dag}|\phi\rangle_3.
\end{eqnarray}
Then, we just need to perform the unitary operation 
${\tilde U}_3^{(0)}$ to complete the teleportation. Eventually, if we
perform again the operations of reading of the ancilla, it is possible to
bring back the ancillas to their initial value
$|0\rangle_{A_1}|0\rangle_{A_2}$, but this is not necessary for the
teleportation. 

\section{Reversible teleportation protocol: generalization to arbitrary number of sites}
\label{sec:gen}

In this section we shall show how our teleportation scheme can be extended to
the case in which we have $N$ sites, instead of $3$ as in Sec.\ \ref{sec:exa}.
We shall consider that $N$ is an even number, i.e. $N=2m$, but similar
procedures can be designed for odd $N$ (as we have already shown for the case
of $N=3$ in Sec.\ \ref{sec:exa}). Therefore, our physical system is now
composed of $N$ two--level atoms each in one site of the lattice, and $2$
ancillas. We consider that the initial state of the system is of the form:
\begin{equation}
\ket{\phi_{in}}=
|0\rangle_{A_1}|0\rangle_{A_2}|\phi\rangle_1\bigotimes_{j=2}^N|0\rangle_j,
\label{iniN}
\end{equation}

\subsection{Creation of the entanglement between $1, 2\dots , N$}

As for the case of $3$ sites, we perform as a first step a Hadamard
transform in the each one of the sites $1,\cdots,N$.
Then, we perform a shift operation (of the lattices of $|0\rangle$ and
$|1\rangle$), and after that we perform another Hadamard transformation on all
the sites. In Appendix \ref{app:proof}, we demonstrate that after applying the
previous three operations the state of the system becomes
(remember that $m=N/2$):
\begin{equation}
V_a\ket{\phi_{in}}=
\frac{1}{2^{m}}|0\rangle_{A_1}|0\rangle_{A_2}\sum_{J=0}^{2^N-1}c_J
|\tilde{\psi}^{(J)}\rangle_{1\cdots N-1}
{\tilde U}_N^{(J)\dag}|\phi\rangle_N,
\label{HLHN}
\end{equation}
where 
\begin{equation}
|\tilde{\psi}^{(J)}\rangle_{1\cdots
N-1}=\bigotimes_{k=1}^{N-1}|a_k^{(J)}\rangle_k,
\end{equation}
with $J=\sum_{k=1}^{N-1}a_k^{(J)}2^{N-1-k}$, and $c_J$ can be $\pm 1$. 
We define:
\begin{eqnarray}
S_e^{(J)}=\sum_{k=1}^{m-1}a_{2k}^{(J)}, \\
S_o^{(J)}=\sum_{k=1}^{m}a_{2k-1}^{(J)},
\end{eqnarray}
which count  the number of $|1\rangle$'s in the even sites (except
$N$) and in the odd sites, respectively. We show  in App.\
\ref{app:proof}, that the following holds 
if $m$ is an even number:
\begin{itemize}
\item if $S_e^{(J)}\text{mod}\, 2=0$, $S_o^{(J)}\text{mod}\, 2=0$, then \\ ${\tilde
U}_N^{(J)}= W^{(0)}\equiv  1+i\sigma_2$.
\item if $S_e^{(J)}\text{mod}\, 2=1$, $S_o^{(J)}\text{mod}\, 2=0$, then \\ ${\tilde
U}_N^{(J)}= W^{(1)}\equiv 1-i\sigma_2$.
\item if $S_e^{(J)}\text{mod}\, 2=0$, $S_o^{(J)}\text{mod}\, 2=1$, then \\ ${\tilde
U}_N^{(J)}= W^{(2)}\equiv \sigma_3+\sigma_1$.
\item if $S_e^{(J)}\text{mod}\, 2=1$, $S_o^{(J)}\text{mod}\, 2=1$, then \\ ${\tilde
U}_N^{(J)}= W^{(3)}\equiv \sigma_3-\sigma_1$.
\end{itemize}
If $m$ is odd the same is valid but one has to interchange
$S_{e,o}^{(J)}\text{mod}\, 2=0\leftrightarrow S_{e,o}^{(J)}\text{mod}\, 2=1$.

\subsection{Reading of $1,\cdots ,N-1$ using the ancillas}

As for the case of just $3$ sites, we first perform a Hadamard transform in
each one of the ancillas (between levels $|0\rangle$ and $|2\rangle$):
\begin{eqnarray}
H_AV_a\ket{\phi_{in}}=
&&\frac{1}{2^{m+1}}(|0\rangle_{A_1}+|2\rangle_{A_1})(|0\rangle_{A_2}+|2\rangle_{A_2})
\nonumber \\
&& \otimes \sum_{J=0}^{2^N-1}c_J|\tilde{\psi}^{(J)}\rangle_{1\cdots N-1}
{\tilde U}_N^{(J)\dag}|\phi\rangle_N.
\end{eqnarray}
Then, we perform a sweep to the right of the lattice of $|2\rangle$
until placing the ancilla $|2\rangle_{A_1}$ in site $N-1$. 
Then, we displace $|2\rangle_{A_1}$ to the left,
in such a way that a sweep operation (\ref{tablesweep}) is performed every $2$
sites beginning in $N-1$, i.e. in the sites $N-1, N-3,\cdots,3,1$.
After that we sweep back to the right the ancilla state $|2\rangle_{A_2}$, in
such a way that a sweep operation (\ref{tablesweep}) is performed every $2$
sites beginning in $N-2$, i.e. in the sites $N-2, N-4,\cdots,4,2$. Finally we
displace the lattice of $|2\rangle$ back to its original position. The state
of the system becomes: 
\begin{eqnarray}
&&\tilde{V_b}H_AV_a\ket{\phi_{in}}=
\frac{1}{2^{m+1}}\sum_{J=0}^{2^N-1}c_J(|0\rangle_{A_1}+(-1)^{S_e^{(J)}\text{mod}\, 2}|2\rangle_{A_1})
\nonumber \\ 
&& (|0\rangle_{A_2}+(-1)^{S_o^{(J)}\text{mod}\, 2}|2\rangle_{A_2}) 
 \otimes
|\tilde{\psi}^{(J)}\rangle_{1\cdots N-1} {\tilde
U}_N^{(J)\dag}|\phi\rangle_N.
\end{eqnarray}
Performing a new Hadamard
transform in both ancillas, the state of the system becomes:
\begin{eqnarray}
&&{V_b}V_a\ket{\phi_{in}}=
\frac{1}{2^m}\sum_{J=0}^{2^N-1}c_J
|S_e^{(J)}\text{mod}\, 2\rangle_{A_1}|S_o^{(J)}\text{mod}\, 2\rangle_{A_2}
\nonumber \\
&& |\tilde{\psi}^{(J)}\rangle_{1\cdots N-1} {\tilde
U}_N^{(J)\dag}|\phi\rangle_N
\end{eqnarray}
with $V_b=H_A\tilde{V_b}H_A$.

\subsection{Teleportation}

As in Sec. \ref{sec:exa}, we are going to use the values of the ancillas
to teleport the state $|\phi\rangle$ into the site $N$. First we
perform a Hadamard transform in the site $N$. Then, the
lattice of $|2\rangle$ is swept to the right until $|2\rangle_{A_2}$ is in
contact with the site $N$ and interacts following the rule of
(\ref{tablesweep}). Let us call this operation $\tilde V_c$. 
After that we perform again a
Hadamard transform in the site $N$. The
effect of these steps is to change the state of the system into
\begin{eqnarray} 
&& \frac{1}{2^m}\sum_{J=0,S_o^{(J)}\text{mod}\, 2=0}^{2^N-1}c'_J
|S_e^{(J)}\text{mod}\, 2\rangle_{A_1}|0\rangle_{A_2} \nonumber \\
&& \otimes |\tilde{\psi}^{(J)}\rangle_{1\cdots N-1}
 W_N^{(0)\dag}|\phi\rangle_N \nonumber \\
&& + \frac{1}{2^m}\sum_{J=0,S_o^{(J)}\text{mod}\, 2=1}^{2^N-1}c'_J
|S_e^{(J)}\text{mod}\, 2\rangle_{A_1}|1\rangle_{A_2} \nonumber \\
&&|\tilde{\psi}^{(J)}\rangle_{1\cdots
N-1}  W_N^{(1)\dag}|\phi\rangle_N,
\end{eqnarray}
where $c'_J$ can be $\pm 1$.
After this, we move further to the right the lattice of $|2\rangle$ until 
$|2\rangle_{A_1}$ enters in contact with the site $N$, interacts with it 
following the rule of (\ref{tablesweep}), and we sweep back the lattice of
$|2\rangle$ to its original position. Let us call this operation $\tilde
V'_c$. After the joint operation $V_c=\tilde V'_c H_A \tilde V_c H_A$, the state
of the system takes the form:
\begin{eqnarray} 
&&V_cV_bV_a\ket{\phi_{in}}= \nonumber \\
&& \frac{1}{2^m}\left [\sum_{J=0}^{2^N-1}c''_J
|S_e^{(J)}\text{mod}\, 2\rangle_{A_1}|S_o^{(J)}\text{mod}\, 2\rangle_{A_2} 
|\tilde{\psi}^{(J)}\rangle_{1\cdots N-1}\right ]\nonumber \\
&&\otimes   W_N^{(0)\dag}|\phi\rangle_N.
\end{eqnarray}
where again $c''_J$ can be $\pm 1$. Then, we just need to perform the
unitary operation  $  W_N^{(0)}$ to complete the teleportation. As in Sec.
\ref{sec:exa}, if we perform again the operations of reading of the ancilla,
it is possible to bring back the ancillas to their initial value
$|0\rangle_{A_1}|0\rangle_{A_2}$, but as previously this is not necessary for
the teleportation.

\section{Conclusions}
\label{sec:conclu}

In this paper we have presented a teleportation scheme which allows to teleport
an atomic state of a two--level atom confined in some site of an optical
lattice to another two--level atom in a distant site of the lattice, in a
reversible way. In order to achieve that, we have used entanglement procedures
recently developped in Refs. \cite{Jaksch99,Briegel99}. We have shown the
similarities and differences of our teleportation scheme in comparison with
other reversible teleportation schemes. A  difference is that we begin
the teleportation process with a  disentangled system, and perform
a shift operation  which entangles all the sites of the lattice;
in particular the particle which possesses initially the state we want to
teletransport, and the particle in which we want to put the state at the end
of the process, are entangled by this operation. We have shown that by using
two other atoms (ancillas), we can teleport the desired state. This is
achieved using basically unitary one--atom operations and sweep operations;
in particular, only two sweeps of the lattice of a third atomic level
$|2\rangle$ are necessary. The process is fully realized via unitary
transformations, and no measurement is performed; therefore, the process is
completely reversible.

Let us make some remarks concerning other aspects of the 
suggested scheme.
The state $|\phi\rangle$ can be initialised in the site $1$ following a procedure
 similar to that which can be employed in the reading step of the teleportation scheme:
(i) We consider an ancilla $A$ sufficiently separated from the rest of the
atoms, and shine with a laser in such a way that a state
$|\phi\rangle_A=a|0\rangle_A+b|2\rangle_A$ is created; (ii) then, we perform a
Hadamard transform in the site $1$ \cite{foot1b}, and perform a sweep
operation between $A$ and $1$; (iii) a Hadamard transform is performed in the
ancilla, and a second sweep operation between $A$ and $1$ is performed. The
result is that the ancilla becomes $|0\rangle_A$, while the site $1$ acquires
the state $|\phi\rangle_1$ as desired. In a similar way, we can put the final
state of the site $N$ into a sufficiently isolated ancilla
and perform a fluorescence experiment by shining with a laser. This allows us
to read the final state, showing that the teleportation has been actually
produced.

We want to  emphasize that the purpose of this paper was to explain a
possible implementation  of a reversible
teleportation scheme in an optical lattice, and {\em not} to
show the most simple way to transport state $\ket{\phi}$ from one lattice site
to another. This could have been achieved more easily by swapping the state of the
first site with the ancilla as described in the paragraph above, 
and then the state of the ancilla with the second site.
The difference between our scheme and simple swapping is that in our case 
the information is never transported to and localised in the ancilla, but
spread over the total state and then localised in the second site, therefore
exhibiting a non-trivial quantum operation. 

Finally, we have used two ancillas, due to the fact that the ancillas are
two--level atoms, and therefore cannot store four different values as
required in the teleportation scheme. The teleportation can be also performed
with just one ancilla, but the scheme becomes more complicated. Basically what
is needed is a three--step process: (i) in a first step the ancilla reads the
even sites of the lattice (as shown in Sec.\ref{sec:gen}), and after that it is
brought to $N$; after this step the possible states of the site $N$ become
two instead of four; (ii) The reading process is repeated, bringing the
ancilla to its original state $|0\rangle$; (iii) then, the ancilla reads the
odd states as in Sec.\ \ref{sec:gen}, and it is brought to $N$; after this
step the possible states of the site $N$ are reduced to just one, and
therefore one just needs to perform a known unitary transformation to conclude the
teleportation.

We hope that the presented teleportation protocol will
motivate further efforts into the realization of simple quantum
networks in optical lattices.

We are grateful to J. I. Cirac for introducing us to this subject, and to 
M. Lewenstein for fruithful discussions.
This work was supported by
Deutsche Forschungsgemeinschaft under grant SFB 407 and by 
the EU through the TMR network ERBXTCT96-0002.

\appendix
\section{Proof of equation (\ref{HLHN})}
\label{app:proof}

In this appendix we  prove that after applying on the initial state
(\ref{iniN}) a Hadamard transform in all the $N$ sites (in the following we
denote this operation as $ H^{\oplus N}$), performing a shift operation (in
the following we call it $ L$), and applying again $ H^{\oplus N}$, 
the state of the system becomes that of expression
(\ref{HLHN}). We are  proving this using induction arguments. Let us call 
${\cal Q}_N=H^{\oplus (N)} L H^{\oplus (N)}$. It is easy to
observe that for the case of $4$ sites, expression (\ref{HLHN}) is fulfilled.
Let us assume that for the case of $N=2m$ sites, (\ref{HLHN}) is fulfilled,
i.e. \begin{equation}
|\Phi\rangle_{1,\cdots,N}=\frac{1}{2^m}\sum_{J=0}^{2^N-1}c_J
|\tilde{\psi}^{(J)}\rangle_{1\cdots N-1}
{\tilde U}_N^{(J)\dag}|\phi\rangle_N.
\end{equation}
with $c_J=\pm 1$. Now, we are going to add two more sites $(N+1),(N+2)$ 
at the right of the 
site $N$. The effect of ${\cal Q}_{N+2}$ on the
initial state for the $N+2$ sites, can be easily calculated from the state
$|\Phi\rangle_{1,\cdots,N}$ using the unitary character of the operations,
then:
\begin{eqnarray}
&&|\Phi\rangle_{1,\cdots,N+2}= \nonumber \\
&& {\cal Q}_{N+2}
\left [ {\cal Q}_{N}^{-1}|\Phi\rangle_{1,\cdots,N}
\otimes |0\rangle_{N+1} |0\rangle_{N+2} \right ] \nonumber \\
&& =\frac{1}{2^{m+1}}\sum_{J=0}^{2^N-1}c_J
|\tilde{\psi}^{(J)}\rangle_{1\cdots N-1}\otimes \nonumber \\
&&\ \ \sum_{k=0}^3 b_k |k\text{div}2\rangle_N|k\text{mod}2\rangle_{N+1} O^{(k)}
{\tilde U}_N^{(J)\dag}|\phi\rangle_{N+2} \nonumber \\
&& =\frac{1}{2^{m+1}}\sum_{J=0}^{2^{N+2}-1}c'_J
|\tilde{\psi '}^{(J)}\rangle_{1\cdots N+1}
{\tilde U}_{N+2}^{\prime (J)\dag}|\phi\rangle_N,
\end{eqnarray}
where $ O^{(0)}\equiv -i\sigma_2$, $ O^{(1)}\equiv
\sigma_1$, $ O^{(2)}\equiv \sigma_3$, 
$ O^{(4)}\equiv  \eins $, and $b_k$ and $c'_J$ can be $\pm 1$. 
Let us assume that $m$ is even, and therefore 
${\tilde U}_N^{(J)\dag}$ satisfies the requirements of Sec.\ \ref{sec:gen}.
It is possible to obtain, after some calculation, that the new unitary
operators ${\tilde U}_{N+2}^{\prime (J)\dag}$ satisfy the same requirements but
interchanging $S_{e,o}^{(J)}\text{mod}\, 2=0\leftrightarrow S_{e,o}^{(J)}\text{mod}\, 2=1$. 
We note that
the same applies if $m$ is odd. Therefore, we have proved that if for $N$
sites (\ref{HLHN}) is satisfied, this  also holds for $N+2$ sites. Since
for $N=4$ the statement is true, the proof is completed by induction.


\begin{references}
\bibitem{foot1b} These one--atom operations can be achieved by shining a laser
on the desired atom. However, it is not realistic to assume that only one
atom is then affected, because the atoms are typically separated by a $\lambda
/2$ distance, where $\lambda$ is the wavelength of the laser which creates the
lattice. Fortunately, our teleportation scheme does not employ one--atom
unitary operations specifically for one site, but operations which can in
principle be applied simultaneously to all the atoms of the lattice. The only
exception is the ancilla, which is assumed to be placed sufficiently 
far away
from the  other atoms of the lattice, and therefore individual unitary
operations can be performed on this particular atom.
\bibitem{foot2} In principle,  kinetic phases appear as well, due to the movement of
the atoms, but these are trivial one--particle phases, which can always be
incorporated into the definition of states $|0\rangle$ and $|1\rangle$.
\bibitem{foot3} In principle,
$|2\rangle_A|0\rangle_j\rightarrow e^{i\phi_0}|2\rangle_A|0\rangle_j$, whereas 
$|2\rangle_A|0\rangle_j\rightarrow e^{i(\phi_0+\phi)}|2\rangle_A|1\rangle_j$,
but as the phase $\phi_0$ appears anyway, it can be reabsorbed in the
definition of $|2\rangle_A$, as it is also done with the kinetic phases
\cite{foot2}. By setting $\phi=\pi$, we retrieve expression (\ref{tablesweep}).
\bibitem{foot4} As pointed out in Ref.\ \cite{Briegel99},
another, perhaps more elegant, way to solve this problem is to consider a three
dimensional lattice, in which the ancilla can be vertically displaced upwards,
moved, and displaced downwards to the desired site of the lattice.
\end{references}
\end{document}